\documentstyle[aps,prd,epsfig]{revtex}
\tighten
\begin{document}

\title{Constraining non standard recombination: A worked example}

\author{Susana J. Landau}
\address{Facultad de Ciencias Astron\'omicas y Geof{\'\i}sicas, 
         U.N.L.P.,Paseo del Bosque S/N, 1900 La Plata,\\ 
         Provincia de Buenos Aires, Argentina; 
         slandau@natura.fcaglp.unlp.edu.ar}

\author{Diego D. Harari}
\address{Departamento de F{\'\i}sica, FCEyN, Universidad de Buenos Aires\\ 
Ciudad Universitaria - Pab. 1, 1428 Buenos Aires, Argentina; harari@df.uba.ar} 

\author{Matias Zaldarriaga}
\address{Institute for Advanced Study, Princeton, NJ 08540; matiasz@ias.edu}
 
\maketitle

\begin{abstract}

We fit the BOOMERANG, MAXIMA and COBE/DMR measurements of the cosmic
microwave background anisotropy in spatially flat cosmological models 
where departures from standard recombination of the primeval plasma are 
parametrized through a change in the fine structure constant $\alpha$ 
compared to its present value. In addition to $\alpha$ we vary the baryon
and dark matter densities, the spectral index of scalar fluctuations, and
the Hubble constant. Within the class of models considered, the lack of a 
prominent second acoustic peak in the measured spectrum can be accomodated 
either by a relatively large baryon density, by a tilt towards the red in 
the spectrum of density fluctuations, or by a delay in the time at which 
neutral hydrogen formed. The ratio between the second and first peak  
decreases by around 25\% either if the baryon density $\Omega_bh^2$ is 
increased or the spectral index $n$ decreased by a comparable amount, or if 
neutral hydrogen formed at a redshift $z_*$ about 15\% smaller than its 
standard value. We find that the present data is best fitted by a delay 
in recombination, with a lower baryon density than the best fit if 
recombination is standard. Our best fit model has $z_*= 900$, 
$\Omega_bh^2=0.024$, $\Omega_mh^2=0.14$, $H_0=49$ and $n=1.02$. 
Compatible with present data at 95\% confidence level $780< z_*<1150$, 
$0.018<\Omega_bh^2<0.036$, $0.07< \Omega_m h^2 < 0.3$ and $0.9<n<1.2$.

\end{abstract}



\section{Introduction}\label{intro}

Measurements of the Cosmic Microwave Background (CMB) anisotropy provide 
information about the physical conditions in the universe right before 
decoupling of matter and radiation, and allow for a precise determination of 
many cosmological parameters. The recently published results of the 
BOOMERANG~\cite{deBernardis} and MAXIMA~\cite{MAXIMA} experiments provide 
strong confirmation of the presence of the first acoustic peak in the 
anisotropy angular power spectrum at the degree scale, which suggests that 
the Universe is very close to spatial flatness. BOOMERANG and MAXIMA also 
provided high quality data at smaller angular scales, which do not reveal 
the presence of a prominent second acoustic peak.

The location, height and width of the first peak is in excellent concordance
with the simplest spatially flat cosmological models 
with a nearly scale invariant spectrum of adiabatic density fluctuations 
(such as those motivated by generic inflationary models) and values of all 
cosmological parameters in good agreement with independent observations. 

The lack of a prominent second acoustic peak requires instead some
sort of departure from the simplest models or most likely values of
some cosmological parameters. Three possibilities 
have been highlighted~\cite{Hu,TZ,Peebles,White,Lange}:
i) the power spectrum of primordial density perturbations is tilted in 
the direction that suppresses small scale fluctuations;
ii) the baryonic matter density is slightly above the upper value
expected from big-bang nucleosynthesis; 
iii) the formation of atomic hydrogen was delayed by some mechanism 
that perturbed the standard ionization history of the Universe.

The first mechanism reduces the height of the second peak simply
because a spectrum tilted towards the red has less power at smaller
angular scales. The second and third mechanisms work because
the ratio between the heights of the second and first peaks decreases 
when the ratio $R=\frac{3}{4}  \rho_ b/\rho_\gamma$
between baryon and radiation energy-densities at
decoupling increases~\cite{Hu95}. Either if $\rho_b$ is increased
or if recombination is shifted to lower redshifts $R(z_*)$ increases and 
the relative height of the second  peak decreases.
The shift of $z_*$ to lower redshifts also shifts the location of
the peaks in the angular power spectrum towards larger angular scales.

Peebles, Seager and Hu~\cite{Peebles} developed a picture where
very early sources of  Ly $\alpha$ resonance radiation   
provide a delay in recombination rapid enough to avoid excessive
dissipation of the first acoustic peak, and cause a 10\% reduction
in the relative height of the second peak along with a 5\% 
shift in the location of the first peak. Their model is essentially 
parametrized by the rate of excess of radiation by the early sources 
from that in the primeval plasma.

In this paper we further explore the compatibility of present CMB anisotropy
measurements with a non standard recombination history. We 
parametrize departures from standard recombination through a change
in the fine structure constant around decoupling compared to its present
value. We then compute constraints in four cosmological parameters
plus the fine structure constant from the COBE/DMR, BOOMERANG and
MAXIMA data on CMB anisotropy. 

In section \ref{alfa} we discuss the effects upon the CMB anisotropy spectrum
of a variation in the fine structure constant at decoupling compared to 
its present value, and argue that they  mimic
the effects of other physical mechanisms that may change the recombination
history. In section \ref{data} we fit the COBE/DMR, BOOMERANG and 
MAXIMA data within 
spatially flat cosmological models allowing for variations of the fine 
structure constant, the baryon and dark matter energy densities, the Hubble 
constant, and the scalar fluctuations spectral index. We place bounds on 
the allowed variation of the fine structure constant, which translate into 
bounds on departures from standard recombination. In section 
\ref{conclusiones} we discuss the results.  

\section{Variation of the fine structure constant as a model of 
non standard recombination}\label{alfa}

Time-dependence in the fine structure constant $\alpha$ modifies the pattern 
of observed cosmic microwave background fluctuations~\cite{Turner,Hannestad}. 
Qualitatively, the main effect of a variation in $\alpha$ 
is due to the change in the binding energy of hydrogen.
A change in $\alpha$ also changes the Thomson scattering cross section and 
modifies the recombination rates.  

We used the CMBFAST code~\cite{cmbfast} with appropriate modifications to 
determine the effects of a change in $\alpha$, along similar lines and with 
comparable results as in \cite{Turner,Hannestad}.

\begin{figure}
\begin{center}
\epsfig{file=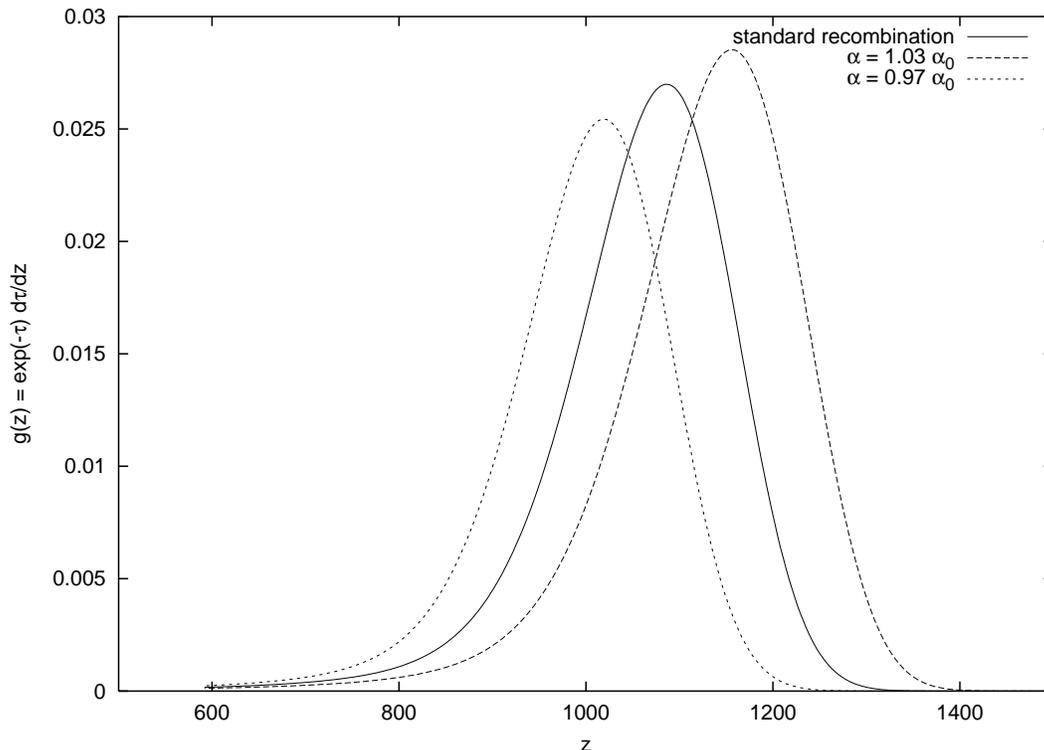,width=4.in,angle=-90}
\caption{The visibility function in a model with standard recombination,
and with recombination delayed (advanced) through a reduction (increase) 
of the fine structure constant $\alpha$ around decoupling by 3\%} 
\label{visi}
\end{center}
\end{figure}

The observed pattern of CMB anisotropies is largely determined by 
the visibility function, $g(z)=exp(-\tau(z)) {d\tau}/dz$, which measures 
the differential probability that a photon last scattered at redshift $z$. 
$\tau$ is the optical depth due to Thomson scattering. Of course, $g(z)$ is
extremely sensitive to the recombination history, since it largely depends
on the time evolution of the fraction of free electrons.

If $\alpha$ was smaller (larger) at recombination than its present value 
$\alpha_0$, the peak in the visibility function shifts towards smaller
(larger) redshifts, and its width slightly increases 
(reduces)~\cite{Turner,Hannestad}. These effects can be appreciated in
Fig.~\ref{visi}, which displays the results of raising and lowering 
$\alpha$ by 3\% around decoupling compared to its present value. 

The hydrogen binding energy scales as $B\propto\alpha^2$. Since the 
ionization fraction is largely determined by the Boltzmann factor
$\exp{(-B/T)}$, the location of the peak of the visibility 
function should roughly scale as $\Delta T_*/T_*\propto 
2 \Delta\alpha/\alpha$ in the limit of small $\Delta\alpha$. 
Although this scaling is not exact~\cite{Turner},
$\Delta z_*/z_*\approx 2 \Delta\alpha/\alpha$ provides a 
reasonably good fit to the shift in the peaks in Fig.~\ref{visi}. 
The width of the visibility function, $\delta z_*$, scales approximately 
as $\Delta (\delta z_*)/\delta z_*\approx -\Delta\alpha/\alpha$. 

Non standard recombination histories should have an effect upon the 
visibility function similar to that described above, except that the 
changes in the peak position and width may follow independent rates of 
change. In this respect, a variation of the fine structure constant alone 
can not mimic an arbitrary mechanism that may delay or advance recombination.
At any rate, a variation of the fine structure constant is a good 
one-parameter emulator of realistic mechanisms for delayed recombination, 
since it incorporates the basic feature of shifting the decoupling redshift 
while maintaining a sufficiently fast recombination rate.
Different physical mechanisms may delay recombination
increasing the width of the last scattering surface at a different
rate, compared to the shift in the peak, than a change in $\alpha$.
The important common feature is that the effect can be differentiated 
from changes in other cosmological parameters, such as the baryon matter
density or the spectral index of density fluctuations.

Indeed, the principal effect of a decrease in the value of the fine structure 
constant around decoupling is due to the shift of $z_*$ to lower values. 
The decrease of $z_*$ boosts the location of the acoustic peaks towards 
larger angular scales. A reduction of $z_*$ also implies an increase in 
$R(z_*)$, the ratio between baryon and photon energy densities at decoupling, 
which increases the height difference between even and odd peaks. 
The height of the first peak is also affected by the change in the 
contribution of modes that entered the horizon during the radiation 
dominated period (the ``early integrated Sachs-Wolfe effect'').  
Finally, the increase in the width of the visibility function that follows
a decrease in the fine structure constant makes diffusion damping more
effective, pushing down the damping tail of CMB anisotropies,
while slightly increasing the still undetected degree of linear 
polarization~\cite{moriond}. This extra diffusion damping may help
differentiate, when higher angular resolution data becomes available,
a decrease of $\alpha$ or a delay in recombination from an increase 
in the baryon density or a red tilt in the spectral index.

Time dependence of the fine structure constant is actually a theoretical
possibility that deserves consideration by its own, not just as a model for 
delayed recombination. Indeed, unification schemes such as superstrings and 
Kaluza-Klein theories predict time variation of fundamental constants over 
cosmological timescales~\cite{Marciano84,DamouryPolyakov,ChodosDetweiler}. 
More recently, a number of authors considered cosmological theories where the 
fine structure constant time dependence is due to the variation of the 
speed of light~\cite{Albrecht99,Barrow99,Clayton99} (constraints on these
theories from CMB anisotropy measurements were analysed in \cite{Avelino00b}). 
Furthermore, 
different versions of the above mentioned theories predict different 
time-dependence of fundamental constants. Thus, experimental bounds on 
their allowed variation are an important tool to check the validity
of these theories.

Constraints on the time variation of the fine structure 
constant have been placed from geophysical and astronomical methods. 
The Oklo natural nuclear reactor
that operated about $1.8\times 10^9$ yrs ago in Oklo, Gabon~\cite{Oklo96}
yields a constraint of $-0.9 \times 10^{-7} <
\Delta\alpha / \alpha<1.2 \times 10^{-7}$. Laboratory 
measurements based on comparisons of rates between clocks with different 
atomic numbers give a limit of $\Delta\alpha / \alpha< 1.4 \times 10^{-14}$ 
during 140 days \cite{Prestage95}. From the analysis of natural 
long-lived $\alpha$ and $\beta $ decayers in geological minerals and 
meteorites Dyson \cite{Dyson66} has placed a bound of  $\Delta\alpha / 
\alpha<2 \times 10^{-5}$. The wealth of local tests, including possible 
correlated synchronous changes of different physical constants, lead to the 
estimate  $\Delta\alpha / \alpha<2 \times 10^{-5}$ for variations during the
last few billion years~\cite{Vucetich90}.

The astronomical methods are based mainly in the analysis of spectra from 
high-redshift quasar absorption systems \cite{Varshalovich00,Webb2,CyS,Webb1}.
Most of the previous mentioned experimental data gave only upper bounds 
(e.g. the most stringent $\Delta \alpha / \alpha=(-4.6 \pm 5.7) 
\times 10^{-5}$ for a set of redshifts $ z \sim 2-4$ \cite{Varshalovich00}). 
Webb et al. \cite{Webb2}, reported a positive measurement of 
the fine structure constant variation: $\Delta \alpha / \alpha=(-1.09\pm 0.36)
\times 10^{-5}$.

Primordial nucleosynthesis provides a constraint to variations in $\alpha$
at the earliest times,
derived form the relative abundance of $^4 He$\cite{Kolb}. However, 
to compute this constraint a model for the $\alpha$ dependence of the
proton to neutron mass ratio must be assumed. L. Bergstr\"om {\it et al.} 
\cite{Iguri99} have arrived a the bound: $\Delta \alpha / \alpha <2\times  
10^{-2}$ including in their analysis not only the $^4He$ abundances but also 
the abundances of other lighter elements that are much less model dependent.

\section{Data Analysis}\label{data}

We have performed a maximum likelihood analysis of the COBE/DMR, 
BOOMERANG and MAXIMA data within spatially-flat
cosmological models with adiabatic density fluctuations. 
We computed the anisotropy correlation multipoles $C_l$ for a grid 
of models allowing for variations in $\alpha / \alpha_0=0.86-1.04,\
\Delta \alpha /\alpha_0=0.02 $;  
$\omega_b=0.01-0.04,\ \Delta \omega_b=0.003$; $\omega_m=0.05-0.6,\
\Delta \omega_m=0.05$; $H_0=45-95,\Delta H_0=5$ ; $n=0.8-1.15,
\Delta n=0.05$.  
Here $\omega_b=\Omega_b h^2,\ \omega_m=\Omega_m h^2$ and $\alpha_0$ is 
the present value of the fine structure constant.
The cosmological constant was fixed to keep each model spatially
flat, so that $\Omega_\Lambda =1- (\omega_b+\omega_m)/h^2$.
The Hubble constant $H_0$ is measured in km/s~Mpc, and
$h=H_0/(100~{\rm km/s~Mpc})$. 
We included calibration errors (20\% for BOOMERANG and 8\% for MAXIMA) 
through the covariance matrix of measurement errors as described 
in~\cite{TZ2}. The amplitude $A_s$ of the scalar fluctuation
spectrum was fixed minimizing $\chi^2$ for each model. 

Our best fit model to the data, with 30 degrees of freedom reduced by 5 
parameters, is characterized by:

\begin{equation}
\omega_b=0.024\ , \ \   \omega_m=0.14  \ , \ \
\frac{\alpha}{\alpha_0}=0.91\ ,\ \
H_0=49  \ ,  \ \ n=1.02  \ , \ \ \chi^2=30.55
\label{BF}
\end{equation}

If we fix as a prior that recombination be standard ($\alpha=\alpha_0$)
the best fit becomes (with now 30 degrees of freedom reduced by 4 
parameters): 

\begin{equation}
\omega_b=0.031\ ,\  \   \omega_m=0.19\ , \ \  
H_0=81  \ , \ \   n=1.02 \ , \  \ \chi^2=33.90
\label{BFPSR}
\end{equation}

If the prior is that the baryonic matter density $\omega_b=0.019$ 
(the nucleosynthesis favoured value~\cite{Tytler}) the best fit is: 

\begin{equation}
\omega_m=0.1 \ , \ \   \frac{\alpha}{\alpha_0}=0.9 \ , \ \
H_0=45 \ , \ \ n=0.96 \ , \ \ \chi^2=34.96
\label{BFPN}
\end{equation}

\begin{figure}
\begin{center}
\epsfig{file=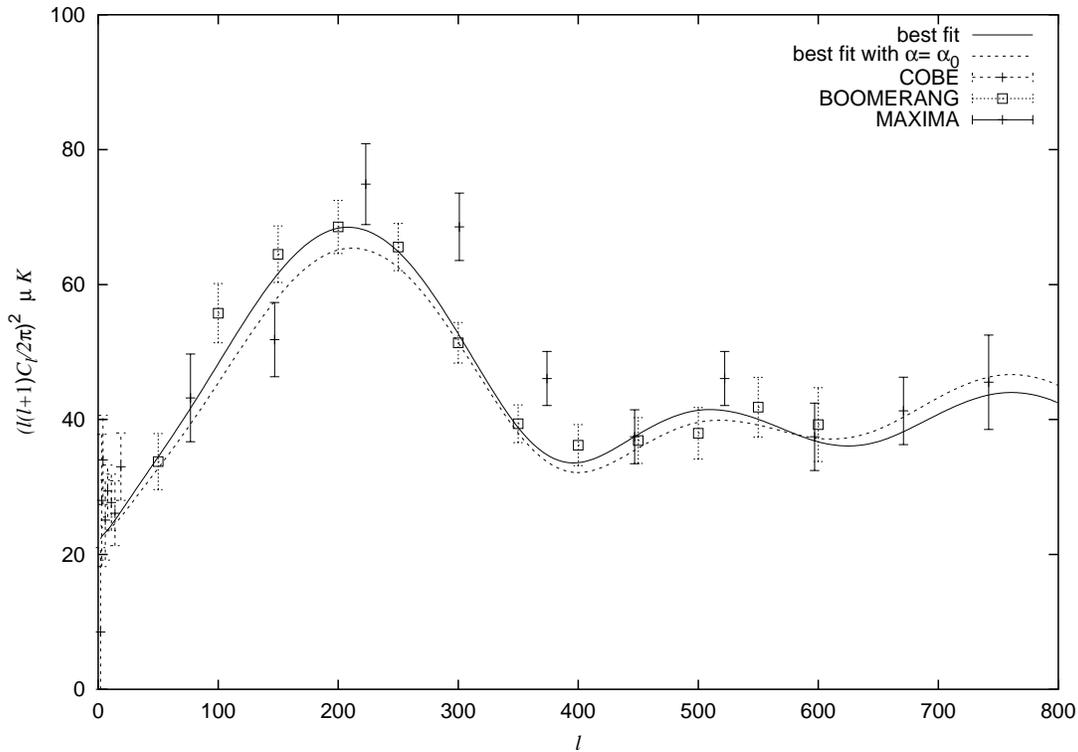,width=4.in,angle=-90}
\caption{Best fit model allowing a variation of the  fine structure constant 
(solid line) and best fit model with standard recombination (dashed line).}
\label{comp}
\end{center}
\end{figure}

Fig.~\ref{comp} displays the anisotropy angular power spectrum for the best 
fit model along with the COBE/DMR, BOOMERANG and MAXIMA data points, as well 
as the best fit model in the case of a standard recombination history.

\newpage

\begin{figure}
\begin{center}
\epsfig{file=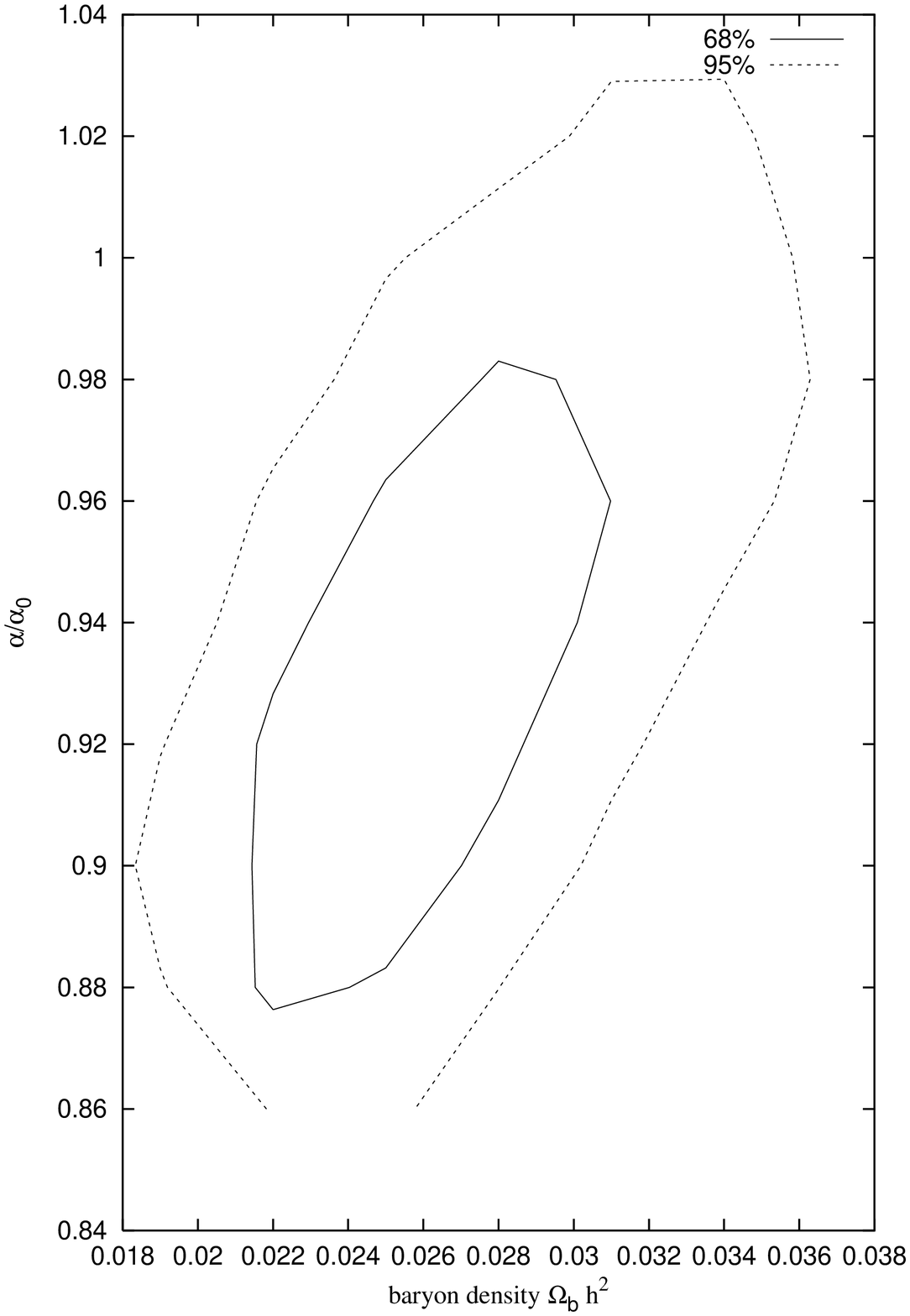,width=3.in}
\epsfig{file=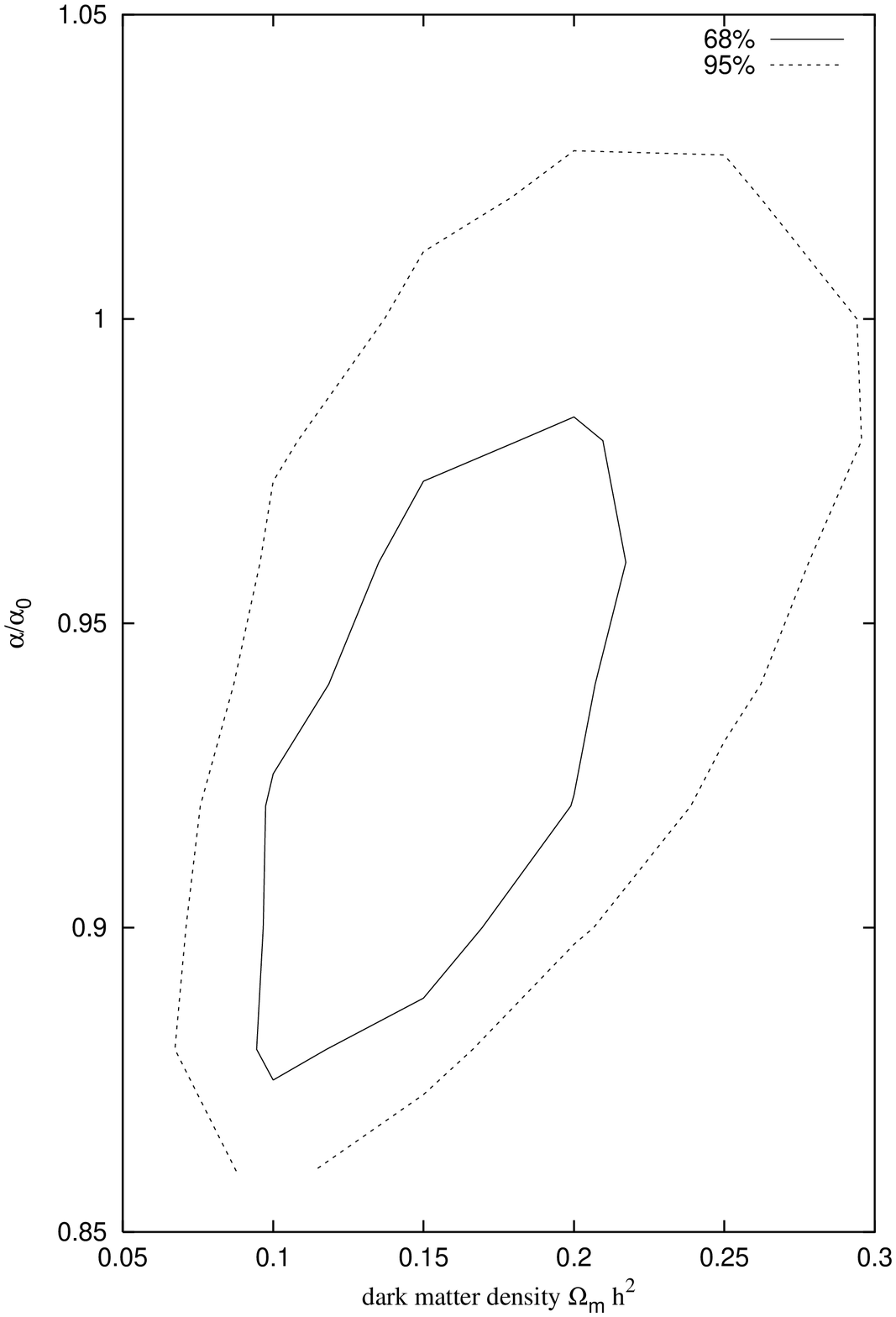,width=3.in}
\vspace{0.25in}
\epsfig{file=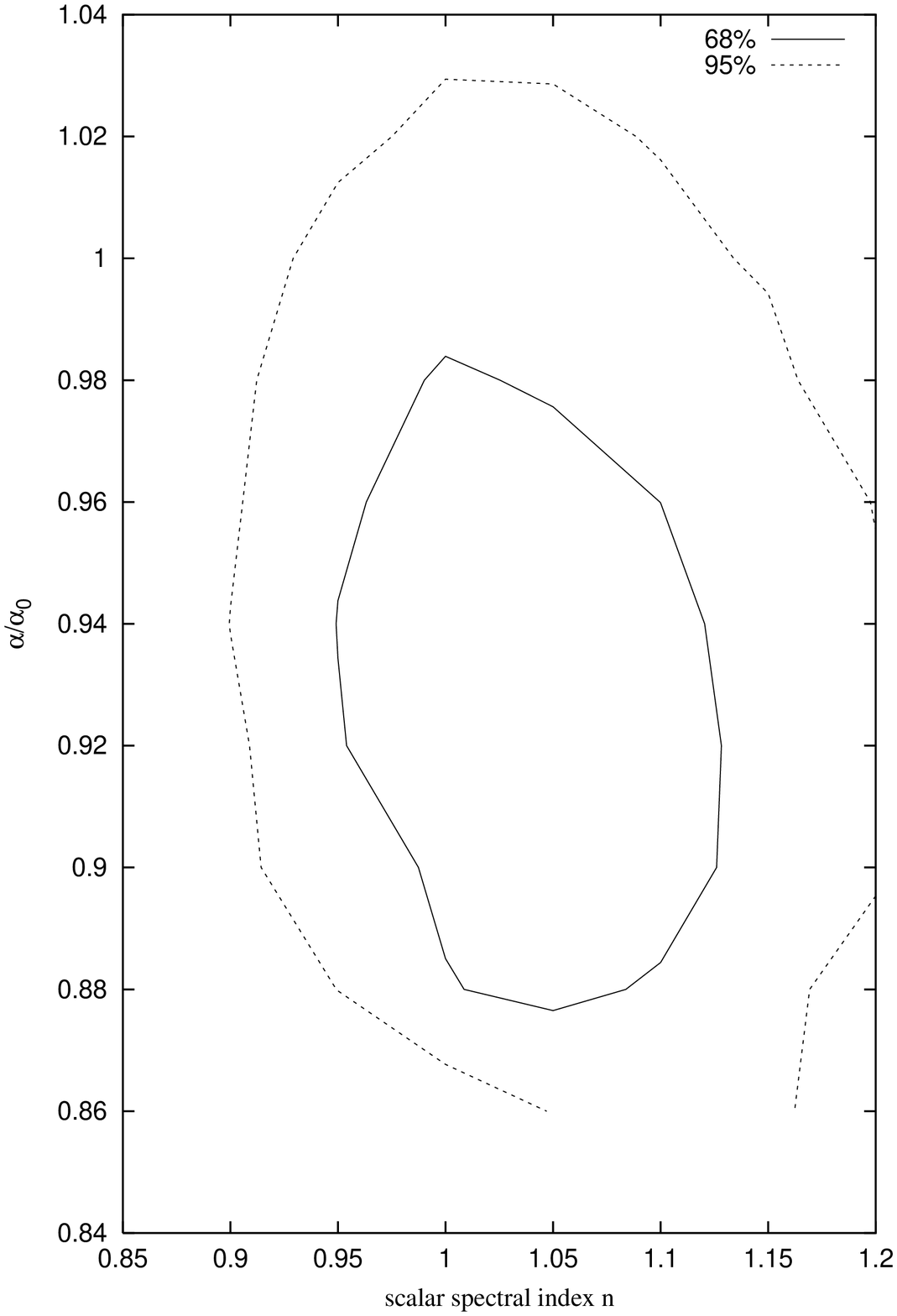,width=3.in}
\epsfig{file=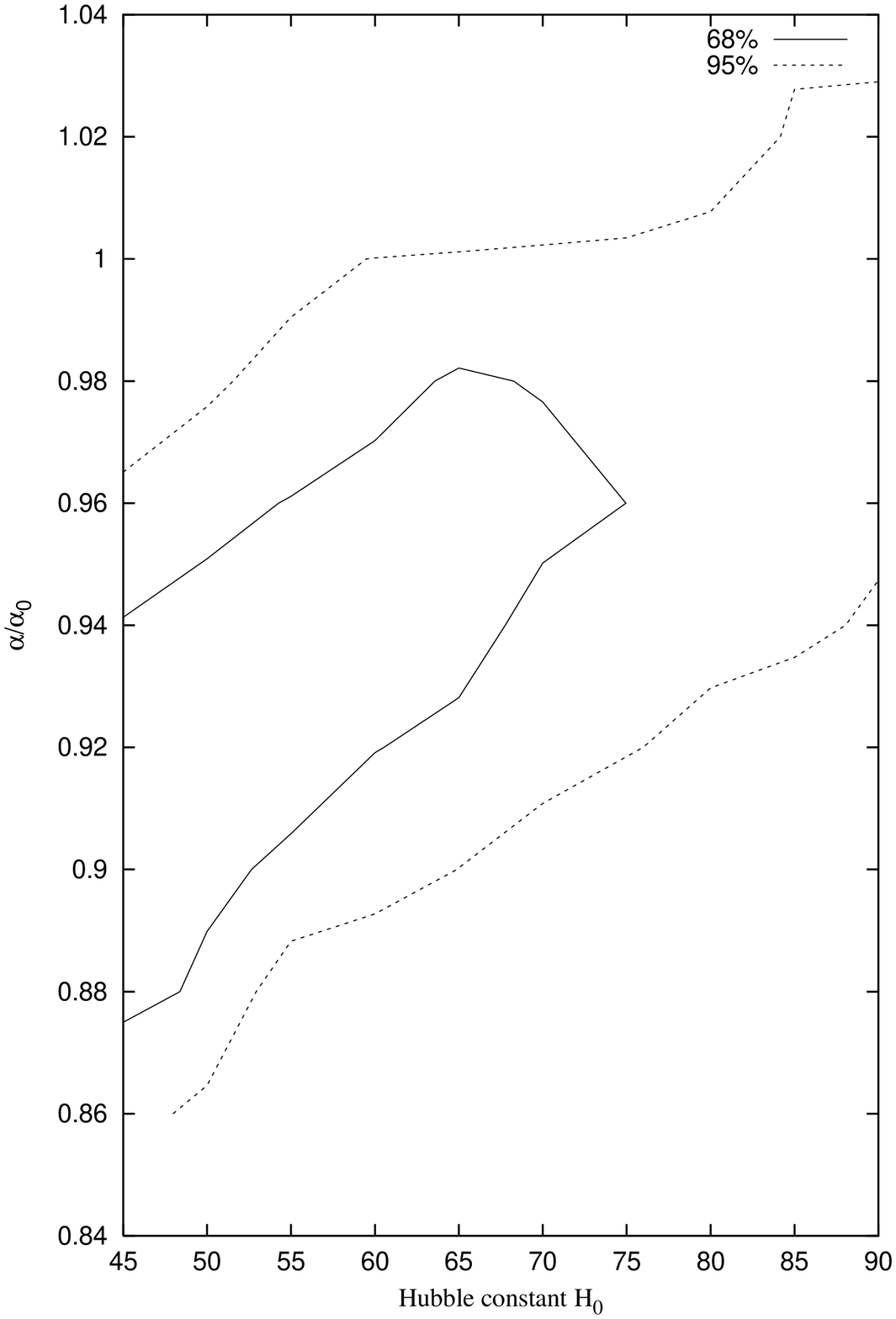,width=3.in}
\end{center}
\caption{1-$\sigma$ and 2-$\sigma$ likelihood contours for 
$\Omega_bh^2$, $\Omega_mh^2$,
$H_0$ and $n$ vs. the relative change in the fine structure
constant in spatially flat cosmological models. The delay in recombination
is $\Delta z_*/z_*\approx 2 (\alpha/\alpha_0 -1)$.}
\label{2D}
\end{figure} 

Two dimensional marginalized likelihood confidence contours of the 
four different cosmological parameters considered as a function of
the variation of the fine structure constant are shown in 
figure \ref{2D}. As we discussed in section \ref{intro}, 
the lack of a prominent
second acoustic peak in the BOOMERANG and MAXIMA data can be accommodated
within simple spatially flat cosmological models either with a
relatively large baryon density, a tilt towards the red in the power 
spectrum, or a delay in recombination. Our contours show that
current data favour delayed recombination over the other options,
which at any rate are all consistent at the 95\% confidence level.

The likelihood contour in the plane ($\omega_b,\alpha$) reveals
a degeneracy between reducing the baryonic matter density
and delaying recombination. The best fit model, in which $\alpha$ 
is smaller by more than 8\% than its present value, and thus
recombination is delayed by around 16\% in redshift, has indeed
a baryon density more than 25\% smaller than the best fit model
with standard recombination. Still, the best fit model has a
baryon density slightly above the nucleosynthesis upper bound.  

The best fit when the baryon density is fixed to the nucleosynthesis
preferred value ($\omega_b=0.019$) requires a large (20\%) delay in 
the recombination redshift, in addition to a small (about 4\%) tilt of
the spectral index towards the red.   

The two-dimensional contours also reveal that fit of the data in
delayed recombination scenarios is best with relatively small
values of the Hubble constant. The degeneracy with the 
value of the spectral index is instead rather small. All the
range allowed for departures from standard recombination at 95\%
confidence level is compatible with a scale invariant ($n=1$) 
spectrum.

\begin{figure}
\begin{center}
\epsfig{file=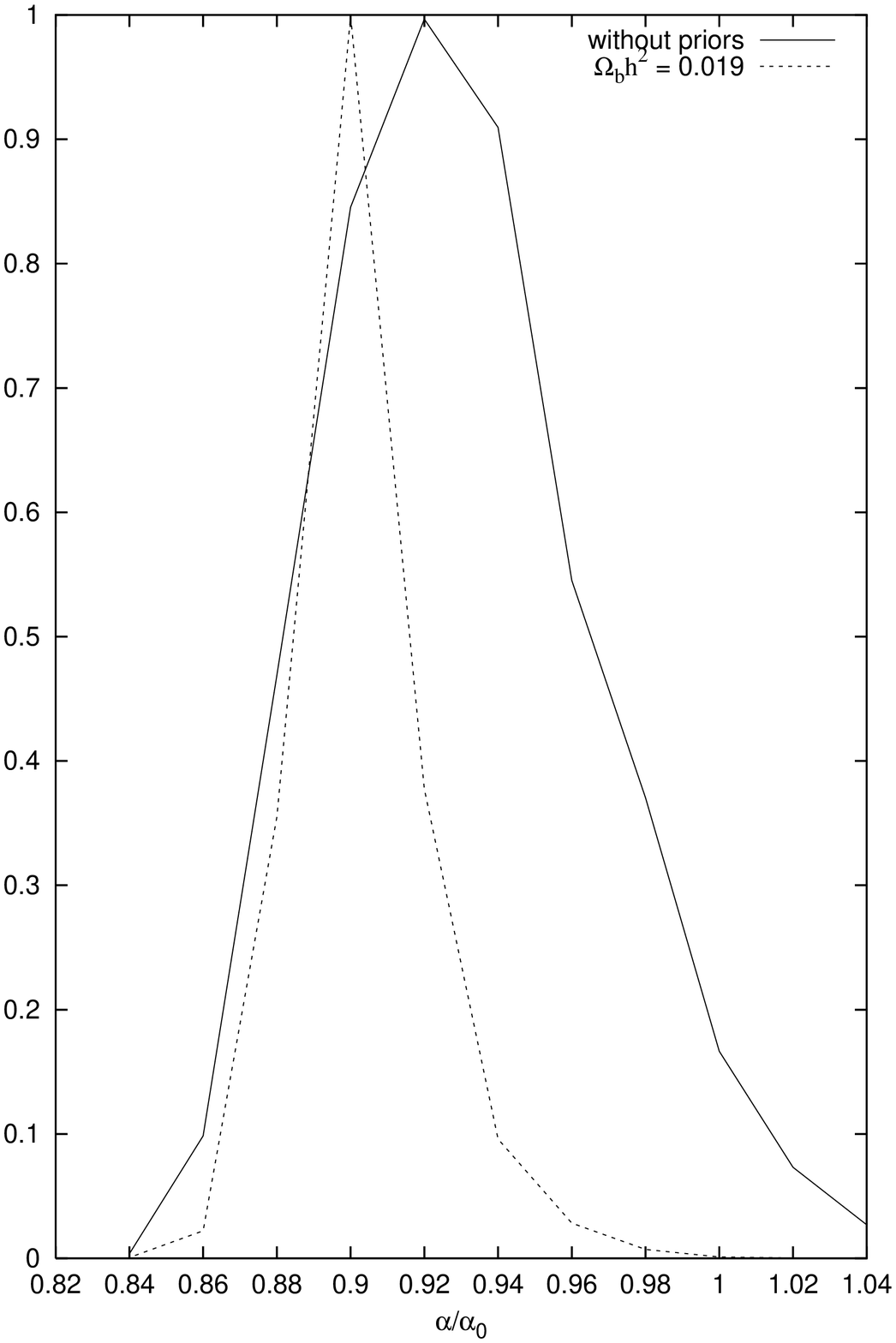,width=3.in}
\epsfig{file=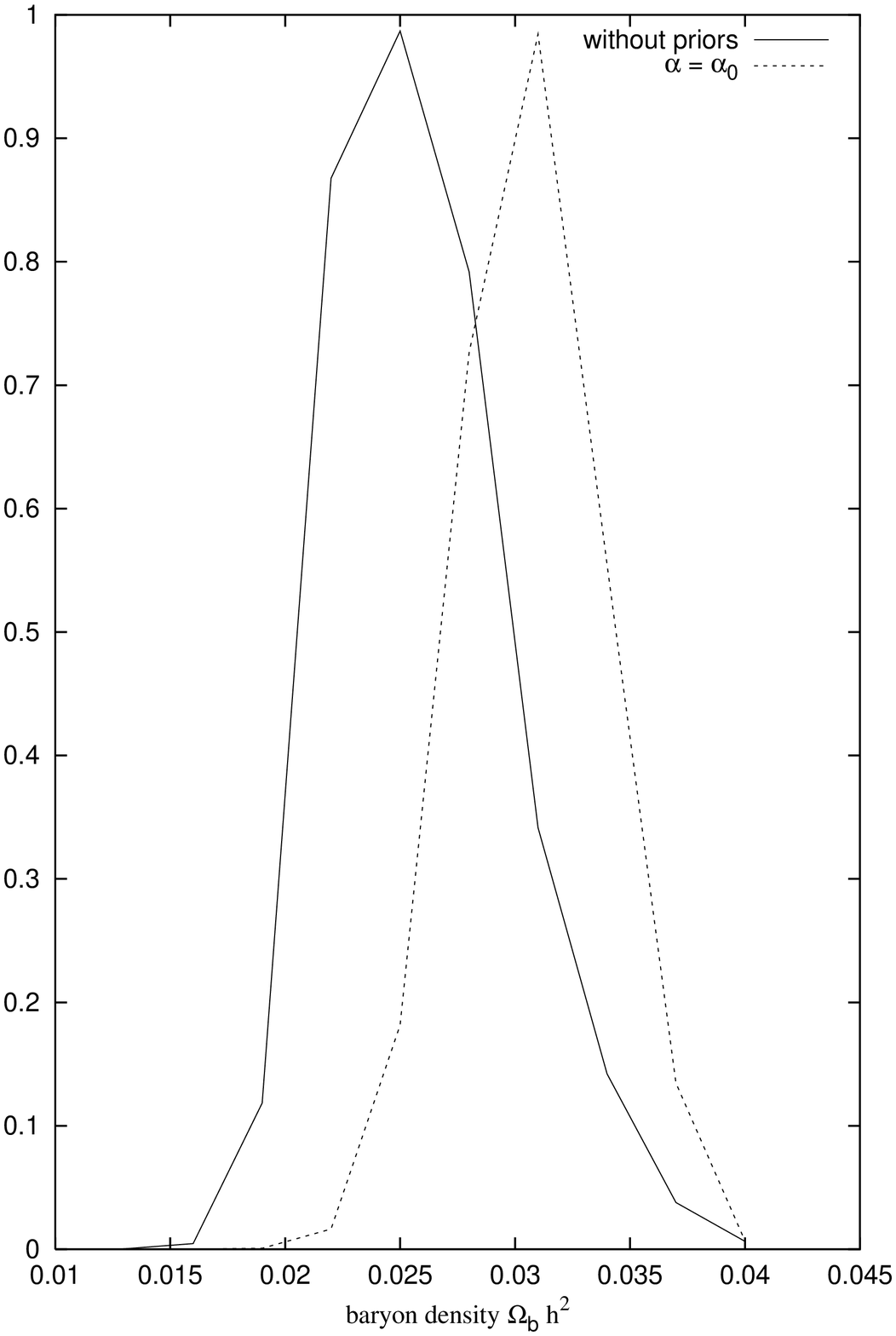,width=3.in}
\end{center}
\caption{One dimensional marginalized likelihoods. Left panel: marginalized 
over $\omega_b,\omega_m, H_0$ and $n$, as a function of $\alpha / \alpha_0$
(without priors) and with $\omega_b$ fixed to its nucleosynthesis prediction;
right panel: marginalized over $\omega_m, H_0$, $n$ and $\alpha$ as a 
function of $\omega_b$ (without priors) and with $\alpha$ fixed to its present
value $\alpha_0$ (standard recombination).}  
\label{1D}
\end{figure} 

One dimensional marginalized likelihoods are shown in figure \ref{1D}.
These make clearer that, within the family of cosmological model considered,
the present data are best fit in a delayed recombination scenario.
If the baryon density is within the nucleosynthesis bound then delayed
recombination is imperative (a tilt in the spectral index alone 
is not enough). 

At 95 \% confidence level, the bounds on the constrained 
parameters are

\begin{equation}
0.018 < \Omega_b h^2 < 0.036\ , \ \ 0.07< \Omega_m h^2 < 0.3\ , \ \ 
0.86 < \alpha / \alpha_0  < 1.03 \ , \ \ H_0 < 95 \ , \ \  0.9 < n < 1.2 
\end{equation}

\begin{figure}
\begin{center}
\epsfig{file=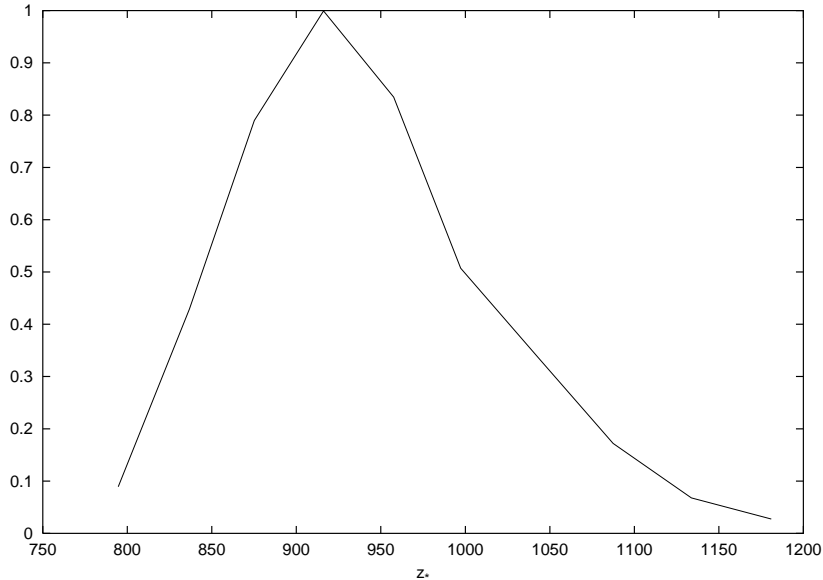,width=3.1in,angle=-90}
\end{center}
\caption{One dimensional marginalized likelihood for $z_*$}  
\label{likez}
\end{figure}

We can use the CMB anisotropy data to place a bound, within the family
of cosmological models considered, upon departures from standard
recombination. We computed $z_*$ (determined as the location of the peak
in the visibility function) for each model in the grid. Then we 
found the model with minimum $\chi^2$ for $z_*$ in given fixed
intervals. From this result we built a one-dimensional likelihood 
for $z_*$, shown in figure \ref{likez}.
    
From figure \ref{likez} we estimate the
allowed range for $z_*$ to be, at the 95\%  confidence level,
\begin{equation}
780<z_*<1150
\label{z*}
\end{equation}
with a best fit value around $z_*=900$.

Notice that the result in eq. \ref{z*} is in very good
agreement with the range in $z_*$ one would expect from the
range $ 0.86<\alpha/\alpha_0<1.03$ and the approximate scaling
$z_*=1080 [1+2(\alpha - \alpha_0)/\alpha_0]$. 

\section{Conclusions}\label{conclusiones}

We have modeled non standard recombination through a variation
of the fine structure constant $\alpha$ around decoupling compared 
to its present value $\alpha_0$.
We performed a maximum likelihood analysis of the BOOMERANG, 
MAXIMA and COBE/DMR measurements of the cosmic microwave background 
anisotropy in a grid of spatially flat cosmological models with five
parameters varied independently: $\alpha$, $\Omega_bh^2$ $\Omega_mh^2$,
$h$ and $n$.  The recombination redshift scales approximately
as $z_*=1080 [1+2(\alpha - \alpha_0)/\alpha_0]$.

The lack of a prominent second acoustic peak in the data
suggests that the best fit should have, within the family of models 
considered, either a relatively large baryon density compared to the
nucleosynthesis value $\omega_b=0.019$, a tilt away from scale 
invariance towards the red in the spectrum of density fluctuations, 
or a delay in the time at which the CMB decoupled from matter. 
The best fit is that recombination was largely delayed
to a redshift around $z_*=900$,
with a nearly scale invariant spectrum of fluctuations, a low value of 
the Hubble constant, $H_0=49$, and a baryon density $\Omega_bh^2=0.024$. 
A delay in recombination allows a good fit to the data with a lower
baryon density as compared to a standard ionization history.
Acceptable fits with a baryon density as low as $\Omega_bh^2=0.018$ can 
be achieved with longer recombination delays.
Good fits with such low baryon densities can not be achieved with just 
a tilt in the spectral index and a standard recombination history.
Recombination can not be delayed beyond $z_*\approx 780$ at 95\% 
confidence level and within the cosmological models considered.
Notice that in our grid of models $H_0>45$. Recombination may eventually
be delayed to lower values of $z_*$ at the same confidence level 
if this (realistic) prior assumption upon $H_0$ were lifted.

It is instructive to discuss these results from a semianalytic approach.
It has been shown~\cite{Hu00} that the power spectrum of CMB anisotropies 
up to the third acoustic peak can be conveniently characterized by 
four observables, namely: the position of the first 
acoustic peak $l_1$, the height of the first peak relative to COBE 
normalization $H_1$, the height of the second peak relative to the first 
$H_2$ and the height of the third peak relative to the first $H_3$. 
We have performed a semianalytic fit around our best fit model
of eq. (\ref{BF}) assuming a linear dependence on the parameters,
with the approximate result:

\begin{eqnarray}
\frac{\Delta l_1}{l_1}&=& 1.4 \frac{\Delta \alpha}{\alpha} + 
0.05 \frac{\Delta \omega_b}{\omega_b} - 0.14 \frac{\Delta \omega_m}{\omega_m}
-0.17  \frac{\Delta H_0}{H_0}+0.19 \Delta n  
\label{l1}\\
\frac{\Delta H_1}{H_1}&=& 0.9 \frac{\Delta \alpha}{\alpha}
+ 0.44\frac{\Delta \omega_b}{\omega_b}
-0.47 \frac{\Delta \omega_m}{\omega_m}-0.13 \frac{\Delta H_0}{H_0}+2.6 
\Delta n
\label{H1}\\
\frac{\Delta H_2}{H_2}&=&3.2 \frac{\Delta \alpha}{\alpha} -
0.85\frac{\Delta \omega_b}{\omega_b}
-0.04 \frac{\Delta \omega_m}{\omega_m}+ 0.96\Delta n 
\label{H2}\\
\frac{\Delta H_3}{H_3}&=&2.3 \frac{\Delta \alpha}{\alpha} -
0.35 \frac{\Delta \omega_b}{\omega_b}+0.38 \frac{\Delta \omega_m}
{\omega_m}+1.2 \Delta n 
\label{H3}
\end{eqnarray}

The peaks location and heights of the best fit model are given
by $l_1=206.5$, $H_1=7.8$, $H_2=0.37$, $H_3=0.43$.

Equation~(\ref{H2}) condenses most of the discussion we already made
in sections~\ref{intro} and~\ref{alfa} about the effects
that can accommodate the absence of a prominent second peak in the present 
CMB anisotropy data.
Indeed, the relative height of the second acoustic peak compared to the
first decreases by about 10\% if $\alpha$ decreases around 3\%. Alternatively, 
the same reduction in $H_2$ can be obtained by an increase of around 10\%
in the baryon density, or a 10\% tilt towards the red in the spectral 
index of density fluctuations. $H_2$ is instead quite insensitive to
$\omega_m$ and $H_0$. 

Notice from eq.~(\ref{l1}) that a  decrease in $\alpha$ by 3\% not only
reduces $H_2$ by around 10\% but also shifts the 
location of the first acoustic peak by almost 5\% towards larger angular 
scales. This coincides with the results
of Ref.~\cite{Peebles}, for a specific model of delayed recombination based
in early Ly $\alpha$ sources of ionizing radiation.

The data from BOOMERANG and MAXIMA does not have the accuracy and 
angular resolution to test for the presence of a third acoustic peak. 
Consequently, we should not use $H_3$ as a relevant parameter. 
Eqs.~(\ref{l1}-\ref{H2}) alone are insufficient to analyse degeneracies as
a function of just one parameter. It is however quite apparent from the
two-dimensional contours of the previous section that the weakest
degeneracy is that on the spectral index $n$, which is quite close
to scale invariance. If we assume $\Delta n=0$ we derive, imposing
$\Delta l_1=\Delta H_1 = \Delta H_2 = 0$, 

\begin{equation}
\frac{\Delta\omega_b}{\omega_b}=3.6~\frac{\Delta\alpha}{\alpha} \ ,
\ \  \frac{\Delta\omega_m}{\omega_m}=3.5~\frac{\Delta\alpha}{\alpha} \ ,
\ \  \frac{\Delta H_0}{H_0}=6.4~\frac{\Delta\alpha}{\alpha} 
\label{deg}
\end{equation}
which is in very good agreement with the shape of the two-dimensional
likelihood contours of the previous section. 

Variations in  $\omega_b$, $\alpha$ and $n$ modify $H_2$.
Neither the variation of $\omega_b$ nor of $n$
have a strong impact upon the location of the first acoustic peak.
Variation of $\alpha$ does, as we already discussed. Compensation
of the peak shift as $\alpha$ decreases  requires a large decrease
either of $H_0$ or of $\omega_m$. Something similar happens with 
$H_1$, which is also very sensitive to changes in the spectral
index $n$. Our results reveal a  strong degeneracy in the plane
($\alpha, H_0$), which appears to be quite robust, in the sense
that it persists with similar strength even under variations of $n$. 
The degeneracy in the plane ($\alpha,\omega_m$) is instead quite 
sensitive to the precise degeneracy in the  $(\alpha,n)$ plane. 
Information on the structure of the angular power spectrum of 
CMB anisotropies around the location of the third acoustic peak 
is necessary to more accurately pinpoint this issue and eventually
break some of the present degeneracies.

A decrease in $\alpha$, as well as any delayed recombination mechanism 
in which there is an increase in the width of the visibility function, 
makes diffusion damping more effective, which increasingly reduces the 
heigths of the peaks at smaller angular scales. 
This is already noticeable around the third acoustic peak. 
A decrease in $\alpha$ that reduces $H_2$ as efficiently as an 
increase in $\omega_b$ has a stronger effect in reducing $H_3$,
as can be seen from eqs.~({\ref{H2},\ref{H3}). 
Notice, for instance, that the best fit model with delayed recombination 
of Fig.~{\ref{comp} has a third accoustic peak suppressed compared to the 
best fit with standard recombination, which has a larger baryon density. 
Measurements of the spectrum around and beyond the third accoustic peak 
can potentially break the degeneracy between the effects of an 
increase in $\omega_b$ and delayed recombination.

Our results can also be seen as a bound on departures from standard
recombination, which can happen at redshifts as low as $z_*=780$ at 
95\% confidence level, or as a bound on the time variation of the
fine structure constant between decoupling and the present time.
The bound we obtain, $0.02>\Delta\alpha/\alpha>-0.14$, is an order of 
magnitude weaker than expected in \cite{Turner} from future satellite 
measurements. The weakness of this bound is a consequence of the
large present degeneracies.
More accurate measurements and higher angular resolutions will certainly
improve this bound. Whether a full order of magnitude improvement is
or is not achieved will depend on how much this degeneracy is 
actually broken.

Our result that there is a preference for a smaller value of 
$\alpha$ at decoupling compared to its present value is in agreement 
with similar conclusions of recent work \cite{Battye00,Avelino00} 
that also analysed BOOMERANG and MAXIMA data allowing for a 
variation in the fine structure constant.
In the present work we used the time varation of the fine structure
constant as an example of how to model non-standard recombination 
histories. A preprint that develops a two-parameter model for 
non-standard recombination histories \cite{Hannestad00} appeared 
after submission of our paper, with compatible results.

\section{Acknowledgements}

S.L. is grateful to Wayne Hu for useful comments.
Support for this work is provided by the Hubble Fellowship (M.Z.) 
HF-01116-01-98A from STScI, operated by AURA, Inc. under NASA 
contract NAS5-26555.
The work of D.H. is supported by grants from ANPCyT and CONICET.
The work of S.L. is supported by the UNLP grant G035.


\begin{thebibliography}{99}

\bibitem{deBernardis}
P. de Bernardis  et al., Nature {\bf 404}, 995 (2000).

\bibitem{MAXIMA} 
S.Hanany et al.,astro-ph/0005123.

\bibitem{Hu}
W. Hu, Nature {\bf 404}, 939 (2000).

\bibitem{TZ}
M. Tegmark and M. Zaldarriaga, Phys. Rev. Lett. {\bf 85}, 2240 (2000).

\bibitem{Peebles}
P. J. E. Peebles, S. Seager and W. Hu, Astrophys.J.{\bf L539}, 1 (2000)

\bibitem{White}
M. White, D. Scott and E. Pierpaoli, astro-ph/0004385.

\bibitem{Lange}
A. E. Lange et al., astro-ph/0005004.

\bibitem{Hu95}
W. Hu and N. Sugiyama, Astrophys. J. {\bf 444}, 489 (1995).

\bibitem{Turner} M. Kaplinghat, R.J. Scherrer and M. Turner, Phys.Rev.
{\bf D 60}, 023516 (1999).

\bibitem{Hannestad}S. Hannestad, Phys.Rev. {\bf D 60}, 023515 (1999).

\bibitem{cmbfast}U. Seljak and M. Zaldarriaga Astrophys. J. {\bf 469}, 437 
(1996).

\bibitem{moriond} S. Landau and D. Harari, in 
Proceedings of the XXXVth Rencontres de Moriond ``Energy densities in the 
Universe'', in press (2000).

\bibitem{Marciano84} W. J. Marciano, Phys. Rev.
Lett. {\bf 52}, 489 (1984).

\bibitem{DamouryPolyakov}  T. Damour and A. M. Polyakov,
Nucl. Phys. {\bf B 423}, 532 (1994).

\bibitem{ChodosDetweiler}  A. Chodos and S. Detweiler,
Phys. Rev. {\bf D 21}, 2167 (1980).

\bibitem{Albrecht99} A. Albrecht and J. Magueijo, Phys.Rev. {\bf D 59}, 043516 
(1999).

\bibitem{Barrow99} J. D. Barrow, Phys. Rev.  {\bf D 59}, 043515
(1999).

\bibitem{Clayton99} M. A. Clayton and J. W. Moffat, Phys. Lett. {\bf B 460}, 
263 (1999).

\bibitem{Avelino00b}P. P. Avelino, C. J. A. Martins and G. Rocha,
Phys. Lett. {\bf B 483}, 210 (2000).

\bibitem{Oklo96} T. Damour and E. Dyson, Nucl. Phys.  {\bf B 480}, 37 (1996).

\bibitem{Prestage95}  D.Prestage, R.L.Toelker and L. Maleki, Phys. Rev. Lett. 
{\bf 74}, 3511 (1995).

\bibitem{Dyson66} F. J. Dyson, Phys. Rev. Lett. {\bf 19}, 1291 (1966).

\bibitem{Vucetich90} P. Sisterna and H. Vucetich; Phys. Rev. 
{\bf D 41}, 1034 (1990).

\bibitem{Varshalovich00}D. A. Varshalovich, A. Y. Pothekin and A. V. 
Invanchik, physics/0004068.

\bibitem{Webb2}  J.K. Webb, V.V.Flambaum, C.W.Churchill, M.J.Drinkwater, and
J.D.Barrow, Phys. Rev. Lett. {\bf 82}, 884 (1999).

\bibitem{CyS}  L. L. Cowie and A. Songaila, Astrophys. J. {\bf 453}, 596 
(1995).

\bibitem{Webb1}  M. J. Drinkwater, J. K. Webb, J. D. Barrow and V. V. Flambaum,
Mon. Not. R. Astron. Soc. {\bf  295}, 452 (1998).

\bibitem{Kolb} E. W. Kolb, M. J. Perry and T. P. Walker, Phys.Rev {\bf D 33}, 
860 (1986).

\bibitem{Iguri99} L. Bergstrom, S. Iguri and H. Rubinstein, Phys.Rev. 
{\bf D 60}, 0405005 (1999).

\bibitem{TZ2} M.Tegmark and M.Zaldarriaga, astro-ph/0002091

\bibitem{Tytler}D. Tytler, J. M. O'Meara, N. Suzuki and D. Lubin, 
Physica Scripta, in press (2000), astro-ph/0001318.

\bibitem{Hu00} W. Hu, M. Fukugita, M. Zaldarriaga and M. Tegmark,
astro-ph/0006436

\bibitem{Battye00}R. A. Battye, R. Crittenden and J. Weller, astro-ph/0008265.

\bibitem{Avelino00}P. P. Avelino, C. J. A. Martins, G. Rocha and P. Viana,
astro-ph/0008446.

\bibitem{Hannestad00}S. Hannestad and R. J. Scherrer, astro-ph/0011188.

\end{thebibliography}
\end{document}